\documentclass[a4paper,twocolumn,nofootinbib,superscriptaddress,aps,prb,eqsecnum,notilepage,showkeys,
]{revtex4-1}
\usepackage{graphicx}
\usepackage{times}
\usepackage{hyperref}
\usepackage[mathlines]{lineno}
\usepackage{natbib} 
\usepackage{amsmath}
\usepackage{amssymb}
\usepackage{setspace}
\usepackage{upgreek}
\usepackage{bm}
\usepackage{mathtools}
\usepackage{ifpdf}
\usepackage{multirow}
\usepackage{gensymb}
\usepackage{mathrsfs}
\usepackage{esvect}
\usepackage{xcolor}
\usepackage{color, soul}
\usepackage[english]{babel}
\usepackage{multirow}
\usepackage{dcolumn}
\usepackage{bbm} 

\begin{document}
\title{High temperature linear magnetoresistance and scaling behavior in the Ba(Fe${_{1-x}}$Co${_{x}}$)$_{2}$As$_{2}$ series}
\author{Rohit Kumar}
\affiliation{Department of Physics, Indian Institute of Science Education and Research \\ Dr. Homi Bhabha Road, Pune, Maharashtra-411008, India}
\author{Surjeet Singh}
\affiliation{Department of Physics, Indian Institute of Science Education and Research \\ Dr. Homi Bhabha Road, Pune, Maharashtra-411008, India}
\affiliation{Centre for Energy Science, Indian Institute of Science Education and Research,\\ Dr Homi Bhabha Road, Pune, Maharashtra-411008, India }
\author{Sunil Nair}
\affiliation{Department of Physics, Indian Institute of Science Education and Research \\ Dr. Homi Bhabha Road, Pune, Maharashtra-411008, India}
\affiliation{Centre for Energy Science, Indian Institute of Science Education and Research,\\ Dr Homi Bhabha Road, Pune, Maharashtra-411008, India }

\date{\today}

\begin{abstract}
	We present magnetotransport studies of the parent, an underdoped and an optimally doped composition of the Ba(Fe${_{1-x}}$Co${_{x}}$)${_{2}}$As${_{2}}$ series. We observe that both the Kohler's and modified Kohler's scaling is typically violated in both the magnetically ordered and paramagnetic regimes. A notable exception is the magnetically ordered state of the underdoped composition where the modified Kohler's scaling is observed, indicating its relative similarity to the cuprates and some heavy fermion systems. This composition also exhibits a feature in the Hall angle, which could signify the opening of a pseudogap before the onset of long range magnetic order. Interestingly, the transverse magnetoresistance is seen to exhibit a linear field dependence in the paramagnetic regimes of all these compositions. We also demonstrate that the $B/T$ scaling proposed recently in the context of quantum critical systems is seen to be valid in all these systems. The implications of our observations are discussed in the context of magnetotransport of metals with incipient magnetic fluctuations.     
		
\end{abstract}

\maketitle

\section{Introduction}

Since the discovery of Iron based superconductors (FeSC),  an increasingly large number of studies have emphasized on the unconventional electronic ground state in these materials. In similarity to other non-BCS superconductors like the cuprates, the heavy fermions, and the organic superconductors, these FeSCs exhibit a number of unique experimental signatures that include an unconventional Cooper pairing symmetry \cite{1}, a strongly temperature dependent Hall coefficient \cite{2}, the possibility of a pseudogap phase \cite{3,4,5}, etc. The differences between these systems are well known - with the cuprates, heavy fermions and FeSCs starting out as being Mott insulators, metals and semimetals respectively. As far as their electronic properties are concerned, the FeSCs are presumably multiband in nature (with multiple electron and hole bands crossing the Fermi level), whereas the single band approximation works to a reasonable extent in the Cuprates and also in some heavy fermions. Prior experimental work has attempted to highlight the commonality between these different classes of superconductors \cite{6,7,8}, and in this context, it is imperative to place the FeSCs in perspective to these other systems. 

Of particular interest here is the exploration of the strange metal phase exhibited by all of these systems, at least in certain regions of their temperature-pressure-doping phase space. Characterized by pronounced non-Fermi liquid like behavior, the normal state magnetotransport is known to exhibit a striking linear temperature dependent resistivity, and unconventional superconductivity is seen to emerge from within this strange metal phase.  For the relatively high temperature superconductors like the cuprates and the FeSCs this has been proposed to signify the presence of a 'Planckian dissipation', where the transport relaxation rate $1/\tau$ is independent of other scattering processes and appears to be driven primarily by the thermal energy scale $k{_B}T$ \cite{zaanen, Bruin}.  

In trying to derive a common thread between material classes as diverse as these, the use of scaling relationships have proven to be invaluable. For instance, the validity of the the quantum critical $\omega/T$ scaling has been demonstrated in the cuprates\cite{9}'\cite{10} and the heavy fermions\cite{11,12,13}, indicating that the physics of these systems are influenced by the presence of a putative quantum critical point (QCP). The Kohler's scaling rule for the magnetoresistance [${\Delta\rho/\rho{_0}} = f(H/\rho{_0})$, where $\Delta\rho$, $\rho{_0}$ and $H$ refer to the magnetoresistance, zero field resistivity and applied magnetic field respectively]  is observed to be satisfied in the (conventional) metallic regions of the cuprates, heavy fermions and FeSCs. The modified Kohler's rule which relates the magnetoresistance to the Hall Angle was first proposed in the context of the cuprates\cite{14}, and has now found utility in evaluating the magnetotransport in the strange metal phase of some heavy fermions\cite{15}'\cite{16} and FeSCs\cite{17}'\cite{18}. Very recently, a new type of scaling, which equates the linear temperature dependence of resistivity with the linear magnetic field dependence of the same quantity ($B/T$ scaling) was proposed\cite{19} in a FeSC near its QCP. This implies that the applied magnetic fields probe the same physics that is accessed by varying the temperature in systems in the vicinity of a QCP. With the validity of this scaling in a heavy fermion system and a cuprate also being demonstrated, it was suggested that this could be a generic signature of the strange metal phase in the vicinity of a QCP.  Intimately coupled to this appears to be the phenomena of linear magnetoresistance - signatures of which have been seen in the FeSCs, and also in some cuprates and heavy fermions (albeit in conjunction with the conventional quadratic contribution)\cite{19,Weickert,Rourke,Butch,Giraldo}.  

In this work, we carried out detailed magnetotransport measurements across the Ba(Fe${_{1-x}}$Co${_{x}}$)${_{2}}$As${_{2}}$ (with $x$ = 0, 0.038 and 0.074) series, with the aim of evaluating the validity of some of these scaling relationships across these systems. In addition, we also report on an anomaly in the Hall angle of the underdoped composition, which reinforces the possibility of a pseudogap in this system - in agreement with earlier transport and spectroscopy measurements\cite{5,shimojima}. An interesting observation is that of linear magnetoresistance, as well as  the validity of the $B/T$ scaling in all these specimens, indicating that this scaling relation also captures the physics of systems  away from the quantum critical region for which it was originally proposed. The rest of the manuscript is organized as follows. Section II deals with crystal growth and experimental details, and Section III provides a brief account of the preliminary characterization of these systems. Section IV presents the experimental results pertaining to the validity (or the lack thereof) of the normal/modified Kohler's scaling for all the three compositions. The observation of linear magnetoresistance in the paramagnetic phase of all three compositions is reported in section V, along with the results of the $B/T$ scaling. Section VI is devoted to discussion of the experimental results, followed by a summary in Section VII.

\section{Materials and Methods}

\section{Preliminary Characterization}

\begin{figure}
	\hspace{-0.5cm}
	\includegraphics[scale = 0.35]{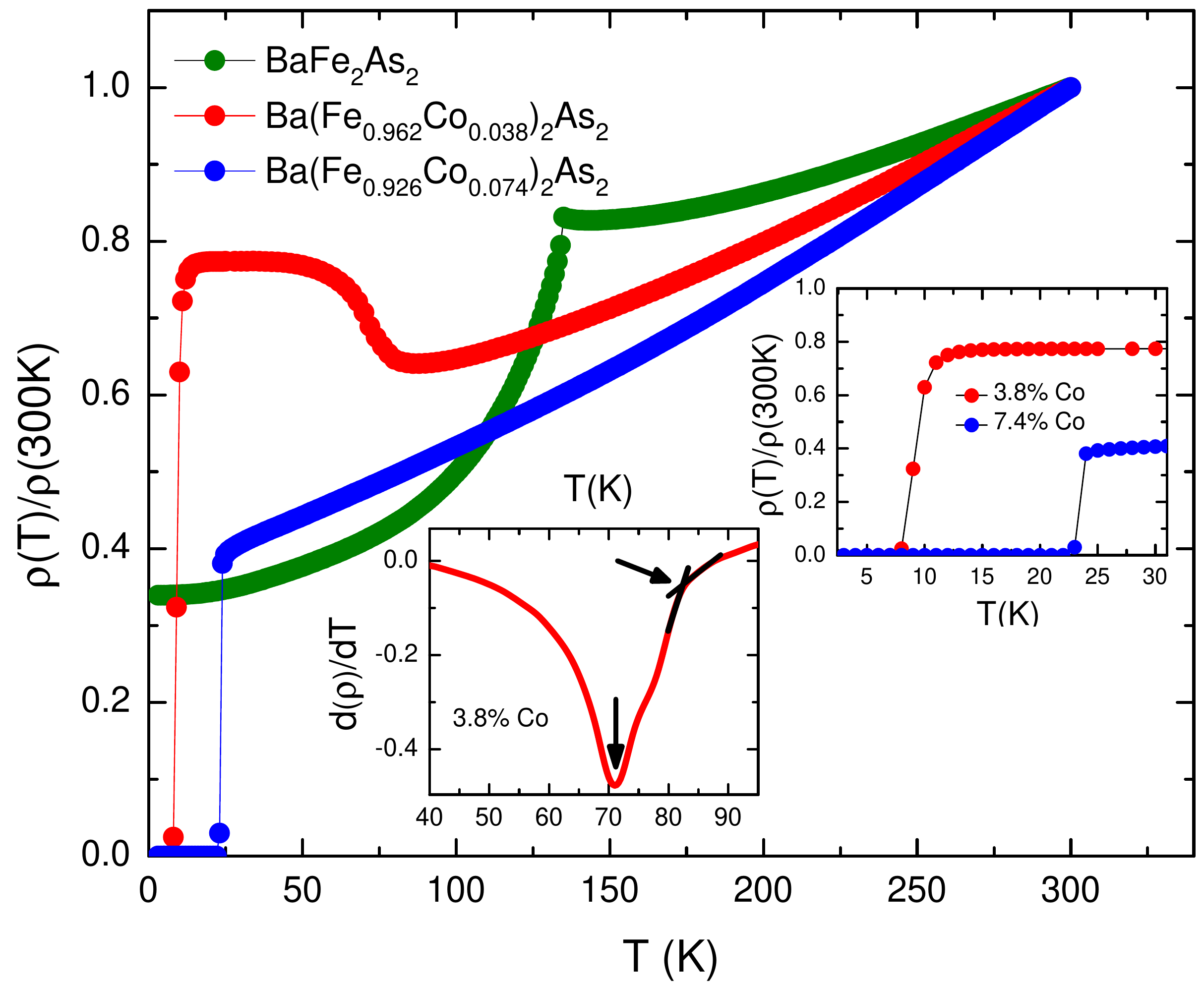}
	\caption{Normalized resistivity for BaFe${_{2}}$As${_{2}}$, the under doped (x = 0.038) and the optimally doped compound (x = 0.074) is plotted. Combined magnetostructural transition in BaFe${_{2}}$As${_{2}}$ is marked by a step like decrease in the resistivity at around 134K. The lower inset shows the structural and magnetic anomalies as observed in the under doped composition. The upper inset depicts an expanded view of the superconducting transitions in the under doped and optimally doped compounds.}
	\label{Figure1}
\end{figure}

As is well known, the parent compound BaFe${_{2}}$As${_{2}}$ goes through combined magnetic and structural transitions around 134 K as is depicted in Figure\ref{Figure1}. Electron doping by replacing Iron with Cobalt, results in the suppression and separation of these transitions, along with the emergence of superconductivity.   In the underdoped composition (x = 0.038), anomalies associated with structural ($T_{S}$) and magnetic ($T_{M}$) transitions can be clearly seen in derivatives of resistivity data (lower inset of Figure\ref{Figure1}) at 83K and 71K respectively, with the onset of superconductivity observed at around 9K. In the optimally doped composition (x = 0.074), the magnetostructural transitions are completely suppressed, and the superconducting transition temperature is observed to be approximately 23.5K (upper inset of Figure \ref{Figure1}).  

\section{Experimental Results}

\begin{figure}
	\hspace{-0.2cm}
	\includegraphics[scale = 0.33]{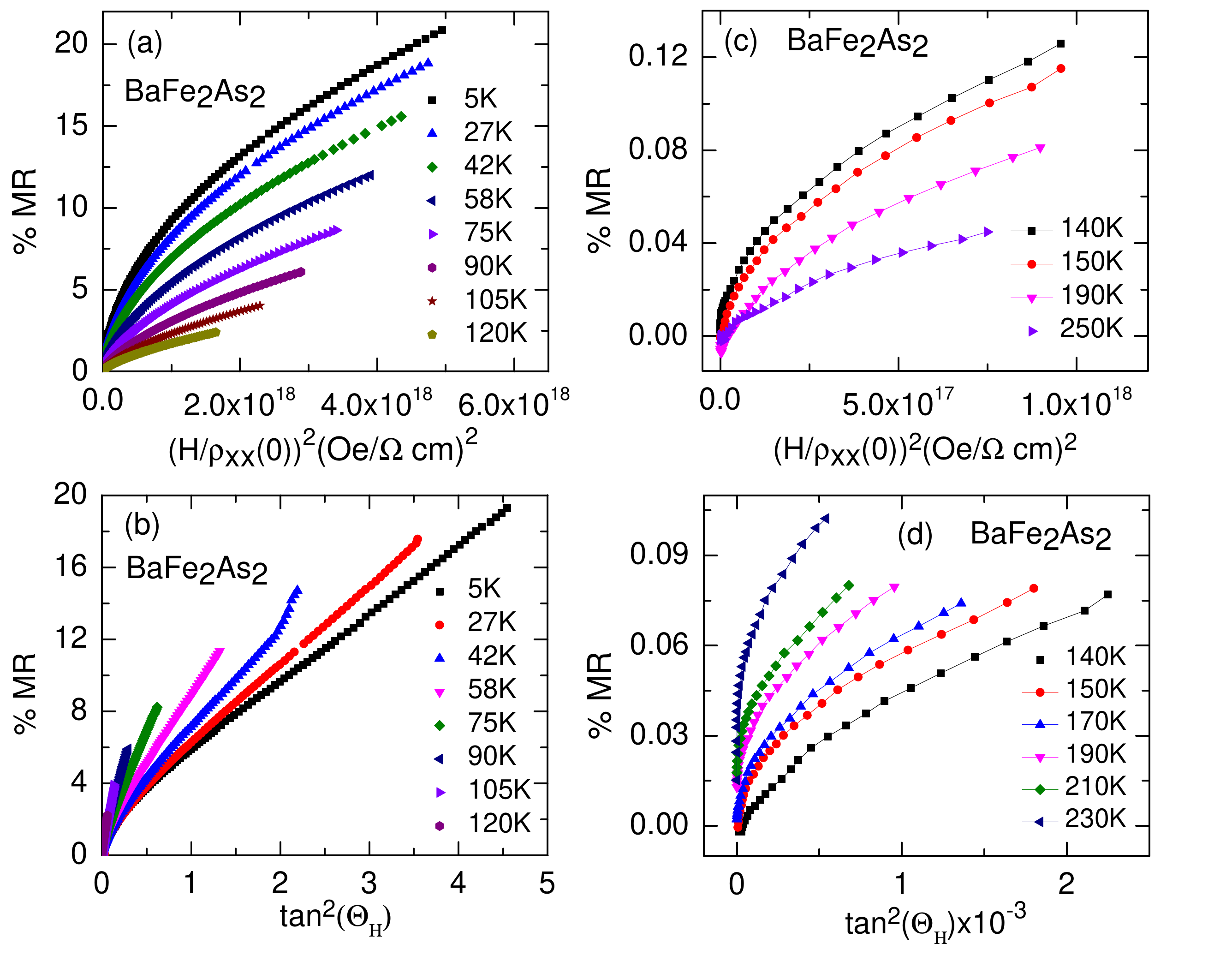}
	\caption{Violation of Kohler's [Figure\ref{Figure2}(a)] as well as modified Kohler's scaling [Figure\ref{Figure2}(b)] in the magnetically ordered state of BaFe${_{2}}$As${_{2}}$ is shown in left panel. Right panel shows violation of Kohler's [Figure\ref{Figure2}(c)] as well as Modified Kohler's scaling [Figure\ref{Figure2}(d)] in the high temperature paramagnetic phase of the same compound.}
	\label{Figure2}
\end{figure}

\subsection{The parent compound BaFe${_{2}}$As${_{2}}$}

The magnetoresistance of many metals can be successfully analyzed using the Kohler's rule which equates the magnetoresistance with the ratio of the applied magnetic field and the zero field resistivity ${\Delta\rho{_{xx}}/\rho{_0}} = f(H/\rho{_0})$, with the temperature being an implicit parameter. In spite of the rather stringent restrictions it places (a single charge carrier along with a uniform scattering rate across the whole Fermi surface), it has found wide utility in a number of strongly correlated systems as well. It has been reported earlier, that Kohler's scaling is violated in the magnetically ordered state of the BaFe${_{2}}$As${_{2}}$ system \cite{22}'\cite{23}, and as is shown in Figure\ref{Figure2}(a), we confirm this observation. This is not surprising, considering the fact that the onset of SDW order would be expected to result in a reconstruction of the Fermi surface, and the violation of Kohler's scaling within a SDW state is known in other systems \cite{24}. It is to be noted that in the case of BaFe${_{2}}$As${_{2}}$, spectroscopy measurements have indicated that there is no opening of a SDW gap, but that the magnetically ordered state is associated with the observation of strong Fermi spots, associated with the emergence of (gapless) Dirac nodes \cite{25,26,27,28}. In the cuprates, the possibility of the reconstruction of the Fermi surface giving rise to two transport times\cite{Schofield, Pines} (preferentially influencing the longitudinal and transverse resistivities) resulted in the reformulation of the Kohler's rule in terms of the Hall angle. This modified Kohler's scaling [${\Delta\rho/\rho{_0}} \propto tan{^2}\theta{_H}$] has been shown to work in the  cuprates\cite{14}, heavy fermions\cite{15}\cite{16}'\cite{8} and in some FeSCs\cite{17}'\cite{18}. However, we observe that the modified Kohler's scaling is also violated in the magnetically ordered state of  BaFe${_{2}}$As${_{2}}$ as is shown in Figure\ref{Figure2}(b). 

Possibly due to very low values of magnetoresistance, such investigations of the paramagnetic state of BaFe${_{2}}$As${_{2}}$ has been relatively scarce. The low magnetoresistance values already indicate that the possibility of the Kohler's scaling being valid in this regime is unlikely, as is also seen in Figure\ref{Figure2}(c).  However, since the presence of quasi-2D spin fluctuations in the paramagnetic regime have been inferred in this compound from prior neutron scattering measurements \cite{29}'\cite{30} it is interesting to check the validity of the modified Kohler's scaling in this regime. It is to be noted that the in cuprates and the heavy fermions, the modified Kohler's scaling is thought to work due to the modification of (the otherwise isotropic) Fermi surface due to short range magnetic fluctuations, with the possible formation of hot spots on the Fermi surface where it intersects the antiferromagnetic Brillouin zone. Interestingly, we observe that the modified Kohler's scaling is also violated in the paramagnetic regime as is shown in Figure\ref{Figure2}(d), clearly indicating that this scenario does not seem to be applicable in the case of BaFe${_{2}}$As${_{2}}$. 

\subsection{The underdoped compound Ba(Fe${_{0.962}}$Co${_{0.038}}$)${_{2}}$As${_{2}}$} 

\begin{figure}
	\hspace{-0.2cm}
	\includegraphics[scale = 0.25]{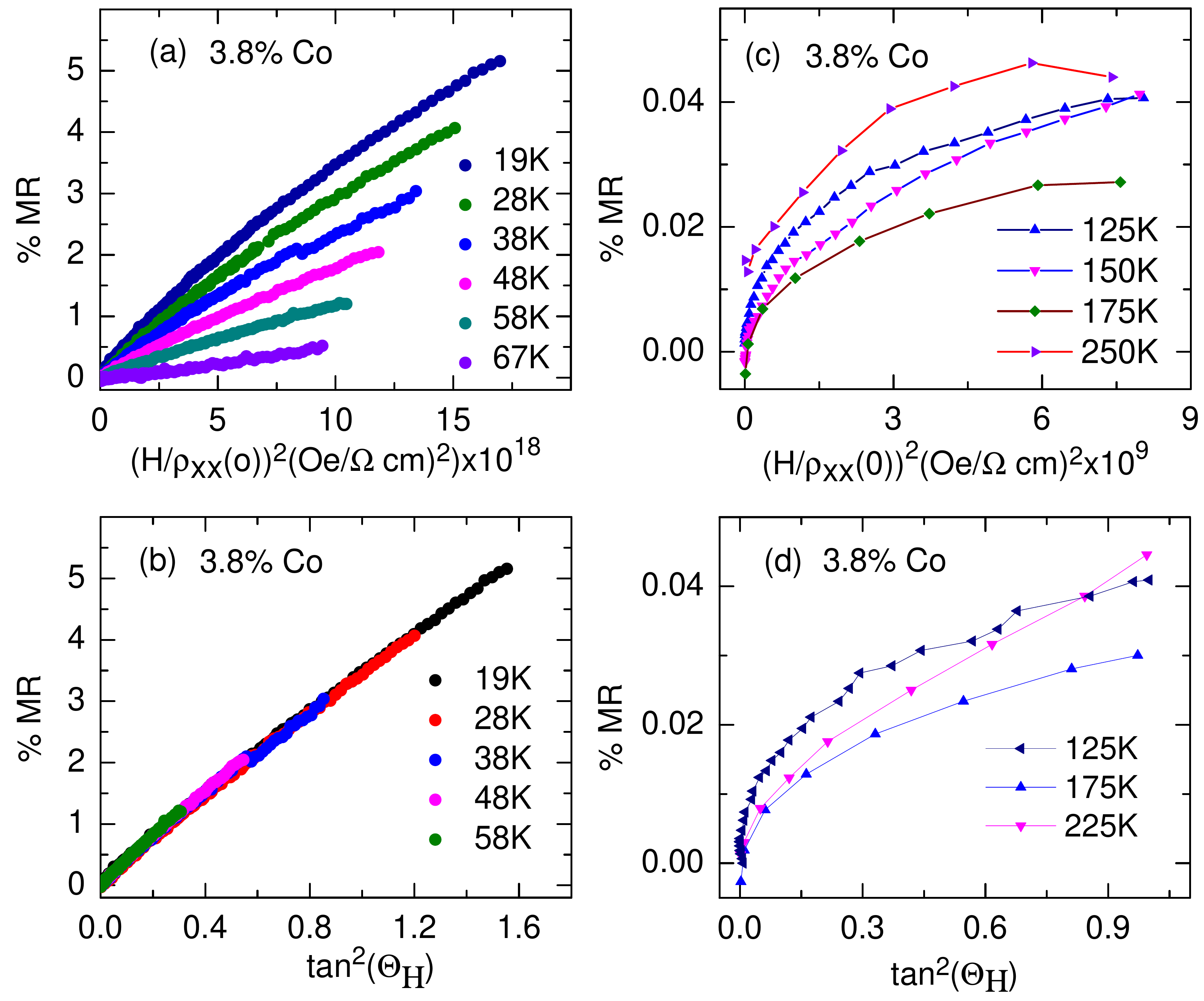}
	\caption{Violation of Kohler's [Figure\ref{Figure3}(a)], and the validity of the Modified Kohler's scaling [Figure\ref{Figure3}(b)] in the magnetically ordered state is shown in the left panel. Right panel depicts the violation of Kohler's[ Figure\ref{Figure3}(c)] as well as Modified Kohler's scaling [Figure\ref{Figure3}(d)] in the high temperature paramagnetic phase.}
	\label{Figure3}
\end{figure}

There has only been a solitary report exploring the feasibility of the Kohler's scaling (or the lack thereof) in any doped member of the Ba(Fe${_{1-x}}$Co${_{x}}$)${_{2}}$As${_{2}}$ family, where superconductivity appears from within a (weakened) SDW state. In an extremely underdoped composition (x = 0.015), it was reported that the Kohler's scaling was violated within the magnetically ordered state, whereas the modified Kohelers scaling appeared to give a better fit\cite{22}. Our observations on the underdoped composition (x = 0.038) are shown in Fig. 3. As is evident, our results are in agreement with this report, with the Kohler's scaling being violated within the magnetically ordered regime (Figure\ref{Figure3}(a)), and the modified Kohler's scaling gives an exceedingly good fit (Figure\ref{Figure3}(b)). We have extended the analysis to the paramagnetic state of this system, where we observe that both the Kohler's (Figure\ref{Figure3}(c))and modified Kohler's (Figure\ref{Figure3}(d)) scaling are violated. In this context, it is interesting to note that the modified Kohler's scaling is observed to work within the paramagnetic state of both the cuprates\cite{14} and the heavy fermions\cite{15}'\cite{16}'\cite{8} and it is evident that the FeSCs clearly do not follow a similar trend. 

The temperature dependence of the Hall angle ($cot\theta{_H} = \rho{_{xx}}/\rho{_{xy}}$) as determined from the isothermal sweeps of the resistivity and the Hall components provides an additional surprise as is shown in Figure 4. We observe a sudden increase in the value of the Hall angle at around 150K, which precedes that of the onset of the SDW, where an additional feature is seen. This feature, which is not immediately discernible from the longitudinal resistivity alone (Fig.1) is seen at temperatures similar to that observed by Tanatar and co-workers using interplane magnetotransport measurements\cite{5}. They had ascribed this to the presence of a pseudogap phase, some evidence of which has also been obtained from Nuclear Magnetic Resonance \cite{4} and Angle Resolved Photo-electron Spectroscopy (ARPES) measurements\cite{31}.  

\begin{figure}
	\centering
	\includegraphics[scale = 0.3]{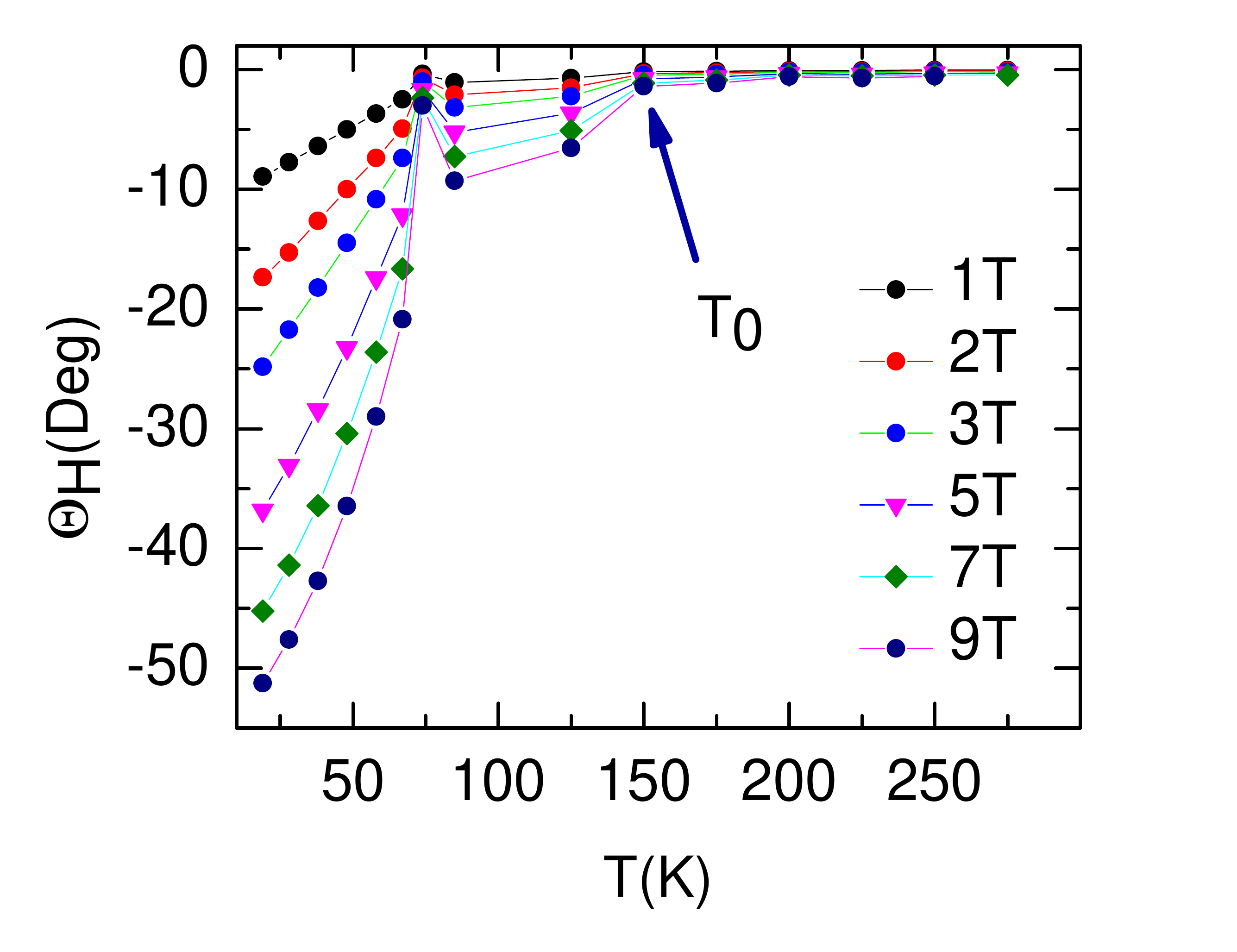}
	\caption{An anomaly associated with the possible formation of a pseudo gap phase(shown by arrow denoted by $T_{0}$) around 150K in the temperature dependent Hall Angle (as measured at different magnetic fields) in the underdoped Ba(Fe${_{0.962}}$Co${_{0.038}}$)${_{2}}$As${_{2}}$ composition.}
	\label{Figure4}
\end{figure}

\subsection{The optimally doped compound Ba(Fe${_{0.926}}$Co${_{0.074}}$)${_{2}}$As${_{2}}$}

The optimally doped compositions refer to those where the doping is equal to (or marginally exceeds the) SDW end point, and superconductivity emerges out of the paramagnetic strange metal phase. Though the cuprates and the heavy fermions, this refers to the region with the most pronounced non-Fermi liquid character, and the validity of the modified Kohler's scaling has been demonstrated in a number of such systems. Though there have been no prior report of the use of such analysis within the electron doped BaFe${_{2}}$As${_{2}}$ systems, there have been sporadic reports of its use in the isovalently substituted members of the BaFe${_{2}}$As${_{2}}$ family. For instance, in the isovalently doped  BaFe${_{2-x}}$Ru${_x}$As${_2}$ system\cite{17},  it has been reported that both the Kohler's and the modified Kohler's scaling is violated for the optimally doped composition, with the modified Kohler's scaling being (partially) recovered only in the overdoped compositions. On the other hand, the optimally doped composition of the isovalently doped  BaFe${_{2}}$(As${_{1-x}}$P${_x}$)${_{2}}$ system\cite{18} exhibits compliance with the modified Kohler's scaling in its paramagnetic region.  Our data indicates that the optimally doped Ba(Fe${_{1-x}}$Co${_{x}}$)${_{2}}$As${_{2}}$ composition is closer to the Ru substituted  BaFe${_{2}}$As${_{2}}$, as both the Kohler's and modified Kohler's scaling does not appear to be valid, as is shown in Figure 5. 

\begin{figure}
	\hspace{-1cm}
	\includegraphics[scale = 0.35]{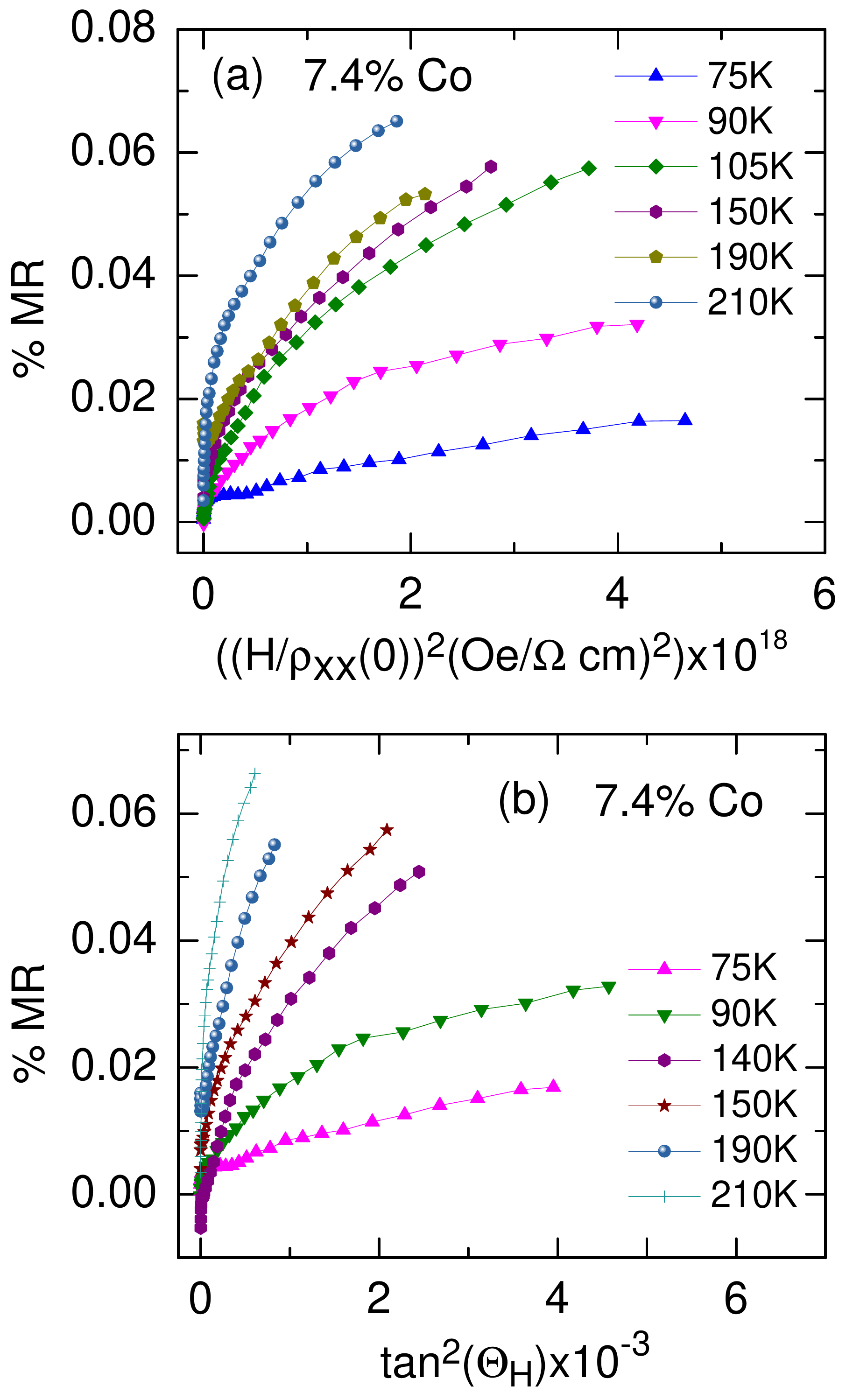}
	\caption{Kohler's plot (a) and Modified Kohler's plot (b) for the optimally doped compound showing clear violations of both scalings in the optimally doped Ba(Fe${_{0.926}}$Co${_{0.074}}$)${_{2}}$As${_{2}}$ system.}
	\label{Figure6}
\end{figure}

\section{Linear Magnetoresistance and $B/T$ scaling}

\begin{figure}
	\hspace{-1cm}
	\includegraphics[scale = 0.35]{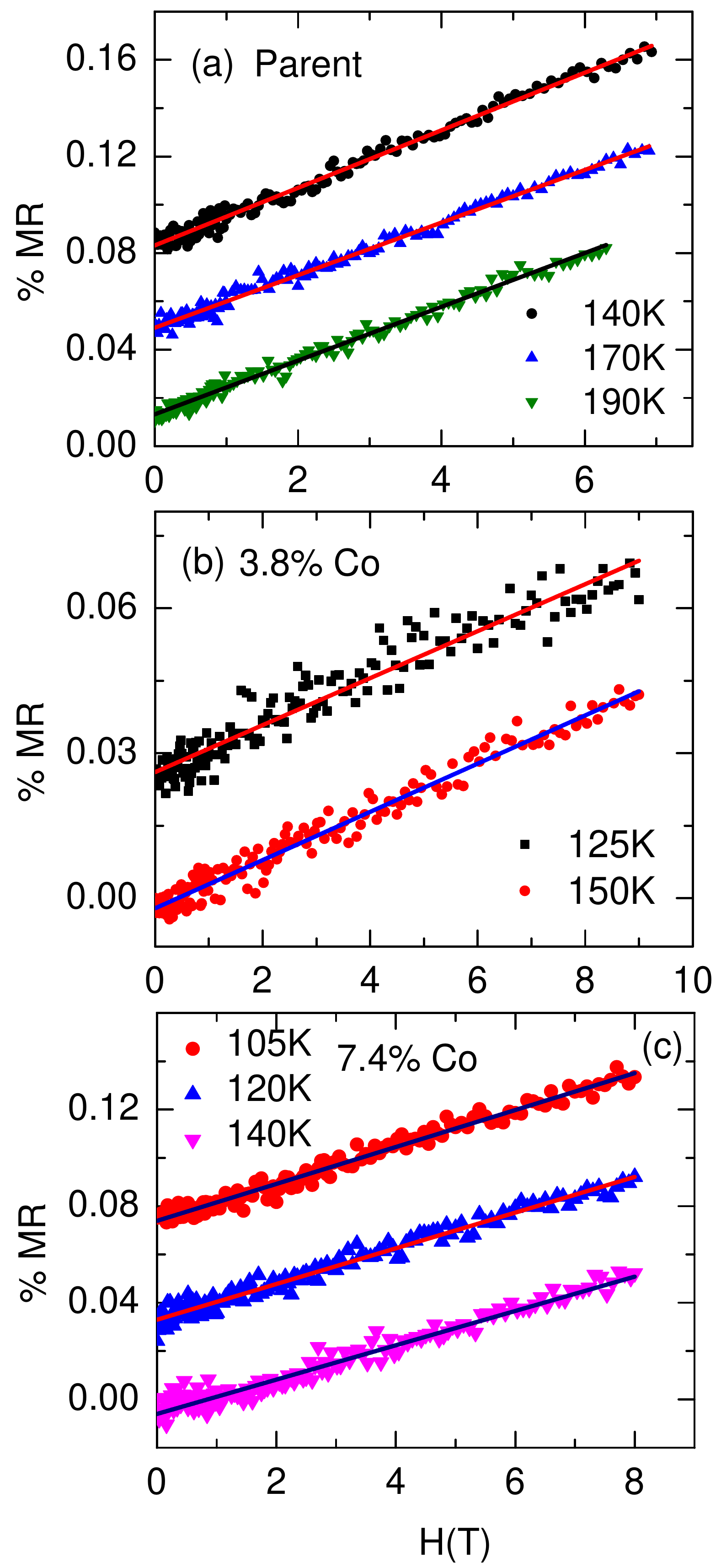}
	\caption{Representative linear Magnetoresistance in the paramagnetic phase of the parent compound BaFe$_{2}$As$_{2}$(a), the under doped compound Ba(Fe$_{0.962}$Co$_{0.038}$)$_{2}$As$_{2}$(b) and the optimally doped compound Ba(Fe$_{0.926}$Co$_{0.074}$)$_{2}$A$_{2}$(c). Plots for each composition are shifted along the y axis for the sake of clarity.}
	\label{Figure7}
\end{figure}

 An interesting manifestation of the unique band topology of the FeSCs has been the observation of a linear $H$ dependence of the transverse magnetoresistance ${\Delta\rho{_{xx}}/\rho{_{xx}}}$\cite{32}. This is in striking contrast to that expected in conventional (compensated) metals, where the high field transverse magnetoresistance would be expected to vary as $H^{2}$ \cite{33}'\cite{34}. First demonstrated in the case of the parent  BaFe${_{2}}$As${_{2}}$, this behavior is now known to exist in doped systems\cite{35}'\cite{22} as well as other members of the FeSC family \cite{36}. This has been attributed to the formation of a Dirac cone state which arises in these FeSC's as a consequence of a special band folding below the antiferromagnetic transition temperature. In the quantum limit, the transverse magnetoresistance\cite{37,38,39} thus varies not as $B^2$, but as $(N{_i}/e{n{_D}}^2)B$, where $N{_i}$ and $n{_D}$ refer to the number of impurities and the number of charge carriers respectively. Since this unique band topology is only expected below the antiferromagnetic transition temperatures, prior reports have primarily concentrated on the low-$T$ magnetically ordered regime. However, on extending the measurements of the transverse magnetoresistance to the paramagnetic regime, we observe that this linear magnetoresistance extends well into the high temperature paramagnetic state, as is depicted in Figure 6, even for relatively moderate fields in excess of 2 Tesla.  
 
\begin{figure*}
	\includegraphics[scale = 0.35]{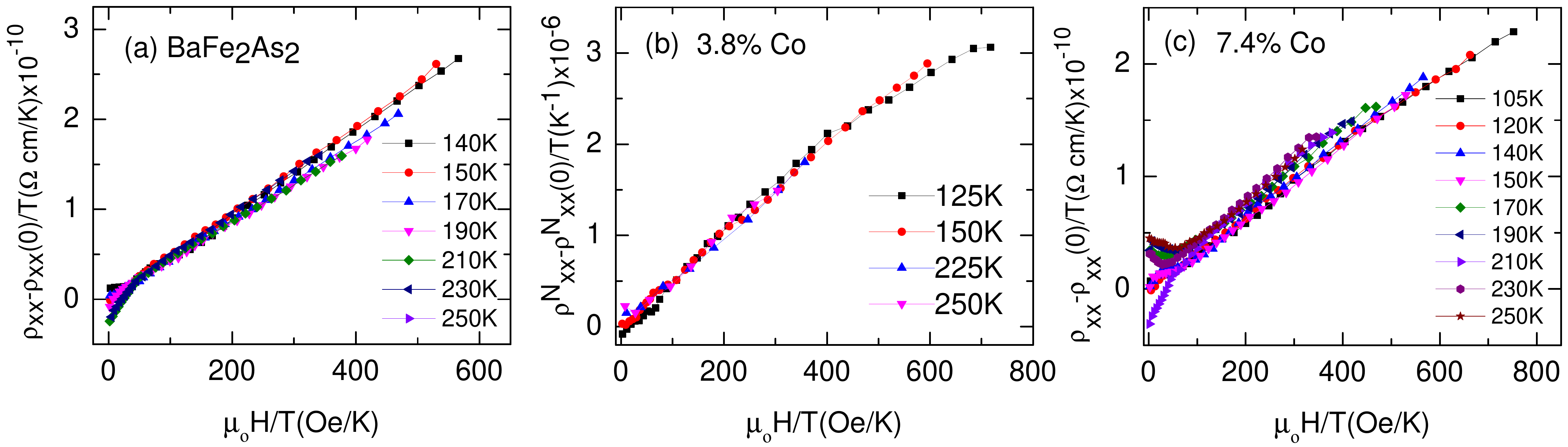}
	\caption{$B/T$ scaling in the paramagnetic region of the parent compound BaFe$_{2}$As$_{2}$(a), the under doped compound Ba(Fe$_{0.962}$Co$_{0.038}$)$_{2}$As$_{2}$(b) and the optimally doped compound Ba(Fe$_{0.926}$Co$_{0.074}$)$_{2}$A$_{2}$(c). Note that some of the plots in (b) are taken from different sets of measurements, and hence the normalized resistivity($\rho_{xx}(H)/\rho_{xx}(8T)$) is used to avoid errors associated with the geometrical factor.}
	\label{Figure8}
\end{figure*}

 The fact that a linear magnetoresistance is observed in the paramagnetic state of both the parent  BaFe$_{2}$As$_{2}$ and the underdoped composition, as well in an extended region of the optimally doped composition (which does not exhibit SDW order) is striking, and could possibly be inconsistent with the Dirac cone scenario invoked to explain the magnetoresistance of a number of such materials. In this context, it is interesting to note that such a linear magnetoresistance has been observed in a number of systems in the vicinity of their quantum critical region \cite{19,Weickert,Rourke,Butch,Giraldo}. It has been suggested that this linear magnetoresistance arises as a consequence of the fact that in quantum critical metals, the magnetic field can set the energy scale of the scattering rate ($\hbar/\tau \sim \mu{_B}B$), in similarity to the temperature. A remarkable manifestation of this similarity was demonstrated for the quantum critical system BaFe${_2}$(As${_{1-x}}$P${_x}$)\cite{19}, where it was demonstrated that the plots of [$\rho(H,T) - \rho(0,0)/T$] vs $B/T$ collapsed on to a single curve. This $B/T$ scaling suggests that the quasiparticle scattering rate $\hbar/\tau$ can be described by a universal function $\sqrt{(\alpha k{_B}T)^{2} + (\eta \mu{_B}B)^{2}}$, with $\alpha$ and $\eta$ being scaling constants that relate the scattering rate to the magnetic field and temperature respectively. Considering the rather moderate magnetic fields used in our measurements, we would expect a larger uncertainty in the estimation of $\rho(0,0)$ used in our analysis. In spite of this, we observe a rather good scaling behavior, as is shown in Figure 7. More surprisingly, unlike what was originally proposed\cite{19}, we observe that this $B/T$ scaling appears to hold true for all the three compositions which we have investigated - at-least in the temperature-field ranges accessed in our measurements.  
 
\section{Discussions}

An important inference from our magnetotransport data pertains to the failure of both the Kohler's and modified Kohler's scaling relationships in describing the magnetotransport of the parent, underdoped, and optimally doped compositions of the Ba(Fe${_{1-x}}$Co${_{x}}$)${_{2}}$As${_{2}}$ series, especially in the paramagnetic state of these systems. A striking exception being that of the magnetically ordered state of the underdoped specimen - where the modified Kohler's scaling appears to be satisfied. This clearly indicates that the low temperature state of the underdoped composition is markedly different from the other regions of the phase diagram, atleast as far as the magnetotransport is concerned. 

As was mentioned earlier, conventional Kohler's scaling relies on the isotropic scattering of a single type of charge carriers across the whole Fermi surface, and a number of possible reasons could result in its breakdown. For instance, the onset of a density wave could result in a reconstruction of the Fermi surface, making the scattering anisotropic along different regions. In some systems, more than one type of charge carriers (with different effective mobilities) could contribute at the Fermi level, thus invalidating the single charge carrier scenario. This is possibly the case in the FeSCs, which are known to be multiband systems, with both electron and hole bands contributing to the Fermi surface, with this multiband character of the Fermi surface  also being a likely prerequisite for superconductivity in these systems. Another possible reason for the violation of Kohler's scaling is the reconstruction of the Fermi surface in terms of 'hot' and 'cold' spots, with disparate scattering rates - a scenario which has been invoked in both the cuprates and the heavy fermions. These different scattering rates (which were thought to preferentially influence the resistivity and Hall conductivities) led to the formulation of the Kohler's scaling in terms of the Hall angle. Our observation that this modified Kohler's scaling is applicable in the low-$T$ state of the underdoped composition indicates that the scenario of a predominantly single carrier type scattering anisotropically at different points in the Fermi surface appears to be valid for this system.  Prior magnetotransport measurements have indicated that the ratio of the hole to electron mobilities decreases sharply as a function of doping, reaching a minimum at concentrations close to that of our underdoped composition \cite{44}. Moreover, it has also been suggested that the underdoped regime is characterized by a competition between spin fluctuations and superconductivity, whereas the overdoped regime suffers from a lack of spin fluctuations - an aspect which was used to explain the asymmetry observed in the superconducting dome of the electron doped  BaFe${_{2}}$As${_{2}}$.  This also offers a potential means of explaining our observations, since the effective electron-like nature of the charge carriers coupled with spin fluctuations in the underdoped composition could mimic the situation commonly observed in the cuprates and some heavy fermions, as a consequence of which the modified Kohler's scaling is followed in this particular region of the phase diagram alone. 

Interestingly, it is in proximity to the same regime that we see signatures of an additional transition, as evidenced from a sharp feature in the Hall angle. The temperature of this downturn matches very well with that reported for a pseudogap phase as determined using interplane magnetoresistance measurements, where a maximum in $\rho{_c}(T)$ was observed. This was also in agreement with earlier Nuclear Magnetic Resonance (NMR) measurements \cite{4}, which also indicated the presence of a pseudugap. This has also now been validated by optical spectroscopy measurements which indicate the presence of a pseudogap, with its spectral manifestations being strikingly similar to that observed in the cuprates \cite{3}. Moreover, it was also suggested that  this gap was intimately coupled with the existence of AFM fluctuations due to the SDW instability, with the gap disappearing at higher doping levels. This further reinforces our observation that in the underdoped regime of Ba(Fe${_{1-x}}$Co${_{x}}$)${_{2}}$As${_{2}}$,  the Fermi surface is strongly renormalized by AFM fluctuations in similarity to the cuprates- as is also evidenced by the apparent validity of the modified Kohler's scaling.   

A surprising observation is that of linear magnetoresistance extending well within the paramagnetic region of all the compositions which we have investigated. As was mentioned earlier, the linear magnetoresistance arising due to Dirac cones is a feature of the magnetically ordered state, where such features arise from complex band foldings due to $d$-band antiferromagnetic interactions. Though the presence of antiferromagnetic fluctuations well into the paramagnetic state has been reported, it is difficult to state with any certainty whether this would be sufficient to enable a topologically induced linear magnetoresistance in this class of materials. We also note that a couple of other alternative theoretical scenarios exist, which could result in similar experimental signatures. For instance, Koshelov \cite{45} proposed a model where the SDW results in the reconstruction of the Fermi surface, near the nesting points, making it difficult for quasiparticles to pass through it during their magnetic field driven orbital motion. As the area of this region varies linearly with the magnetic field, this model also provides a means of explaining the linear magnetoresistance observed in the FeSCs. Moreover a rough estimate of the critical field above which this linearity would be expected was estimated to be $\approx$ 2T, which is also in good agreement with experimental observations. Another relevant model is the one for nearly antiferromagnetic metals in the vicinity of a Quantum Phase transition, where magnetic fluctuations give rise to hot spots on the Fermi surface\cite{46} interrupting orbital motion, and predicting a resistivity linearly varying with $B$. 
 
Closely related to the observation of linear magnetoresistance is the apparent validity of the $B/T$ scaling in all the systems which we have investigated. It is to be noted that this scaling was proposed (and demonstrated) for quantum critical systems alone, and our observation implies that its validity could extend to a regime wider than that for which it  was originally proposed. Interestingly, the failure of this scaling in the overdoped specimens of the the BaFe${_2}$(As${_{1-x}}$P${_x}$) was used to indicate that this scaling phenomena would breakdown away from optimal doping\cite{19}. However, as has been mentioned earlier, the overdoped regime is characterized by much smaller spin fluctuations than the underdoped regime - a factor which could be responsible for the differences observed in the magnetotransport. In light of our observations, it would be interesting to investigate whether this $B/T$ scaling is also valid in the underdoped regimes of other members of the FeSC family.

\section{Conclusions}   

In summary, we report on the scaling analysis of the magnetotransport as measured in three specimens of the Ba(Fe${_{1-x}}$Co${_{x}}$)${_{2}}$As${_{2}}$ series. The Kohler's and modified Kohler's scaling is observed to be violated in both the magnetically ordered and paramagnetic regimes of these specimens, with the notable exception of the underdoped composition, where the modified Kohler's scaling is seen to be valid in the low-$T$ magnetically ordered state. The underdoped specimen also exhibits features similar to a pseudogap, as determined by a sharp feature in the Hall angle. We observe that the paramagnetic regimes of all these compositions exhibit a transverse magnetoresistance varying linearly with the applied magnetic field. A $B/T$ scaling procedure proposed to be valid for quantum critical systems is seen to be be applicable to all these systems, indicating that the correspondence between temperature and magnetic field in dictating the quasiparticle scattering rate could be applicable in a much wider region of the phase space of these materials.   

\section{Acknowledgements}

Rohit Kumar would like to thank Luminita Harnagea for insightful discussions at several occasions regarding single crystal growth of these systems. SS thanks DST-SERB, Govt. of India for financial support under grant no. SR/FTP/PS-037/2010.

\section{References}

\bibliographystyle{apsrev4-1}
\bibliography{Seconddraft}

\end{document}